\documentclass[twocolumn,aps,showpacs,amsmath,amssymbl]{revtex4}
\usepackage{amsmath,bm,epsfig}
\usepackage{color}
\usepackage[normalem]{ulem}

\def\Fbox#1{\vskip1ex\hbox to 8.5cm{\hfil\fboxsep0.3cm\fbox{%
  \parbox{8.0cm}{#1}}\hfil}\vskip1ex\noindent}  

\newcommand{\B}[1]{{\bm{#1}}}
\let \= \equiv \let\*\cdot \let\~\widetilde \let\-\overline

\begin{document}
\title{Percolating Plastic Failure as a Mechanism for Shear Softening in Amorphous Solids}
\author{Vijayakumar Chikkadi$^1$, Oleg Gendelman$^2$, Valery Ilyin$^1$, Ashwin J$^1$ and Itamar Procaccia$^1$}
\affiliation{$^1$Department of Chemical Physics, The Weizmann
 Institute of Science, Rehovot 76100, Israel\\ $^2$ Faculty of Mechanical Engineering, Technion, Haifa 32000, Israel}
\date{\today}
\begin{abstract}
``Shear softening" refers to the observed reduction in shear modulus when the stress on an amorphous solid is increased
beyond the initial linear region. Careful numerical quasi-static simulations reveal an intimate relation between plastic failure and shear softening. The attaintment of the steady-state value of the shear modulus associated with plastic flow is identified with a percolation of the regions that underwent a plastic event. We present an elementary ``two-state" model that interpolates between failed and virgin regions and provides a simple and effective characterization of the shear softening.

\end{abstract}

\maketitle

 ``Shear softening" is a term used in material science to describe the well-known phenomenon of the reduction in the shear or bulk modulus of a given material when the strain (and therefore the stress) in the material increases \cite{07LL,07LC}. Usually one attempts to describe it in terms of nonlinear elasticity disregarding plastic effect \cite{01FO}. Our simulations indicate that a completely different approach may be required.  The phenomenon of shear softening depends on many parameters, like the protocol of preparation of the material, on the temperature, on the strain rate and on the type of loading. In this Letter we aim to simplify the phenomenon to its bare bones, to attain the simplest possible model to explain it. We therefore study 2-dimensional models of amorphous solids which are strained quasi-statically in athermal conditions. This will expose the fundamental physics of shear softening in a way that can be fully understood.

\begin{figure}
\includegraphics[scale = 0.2]{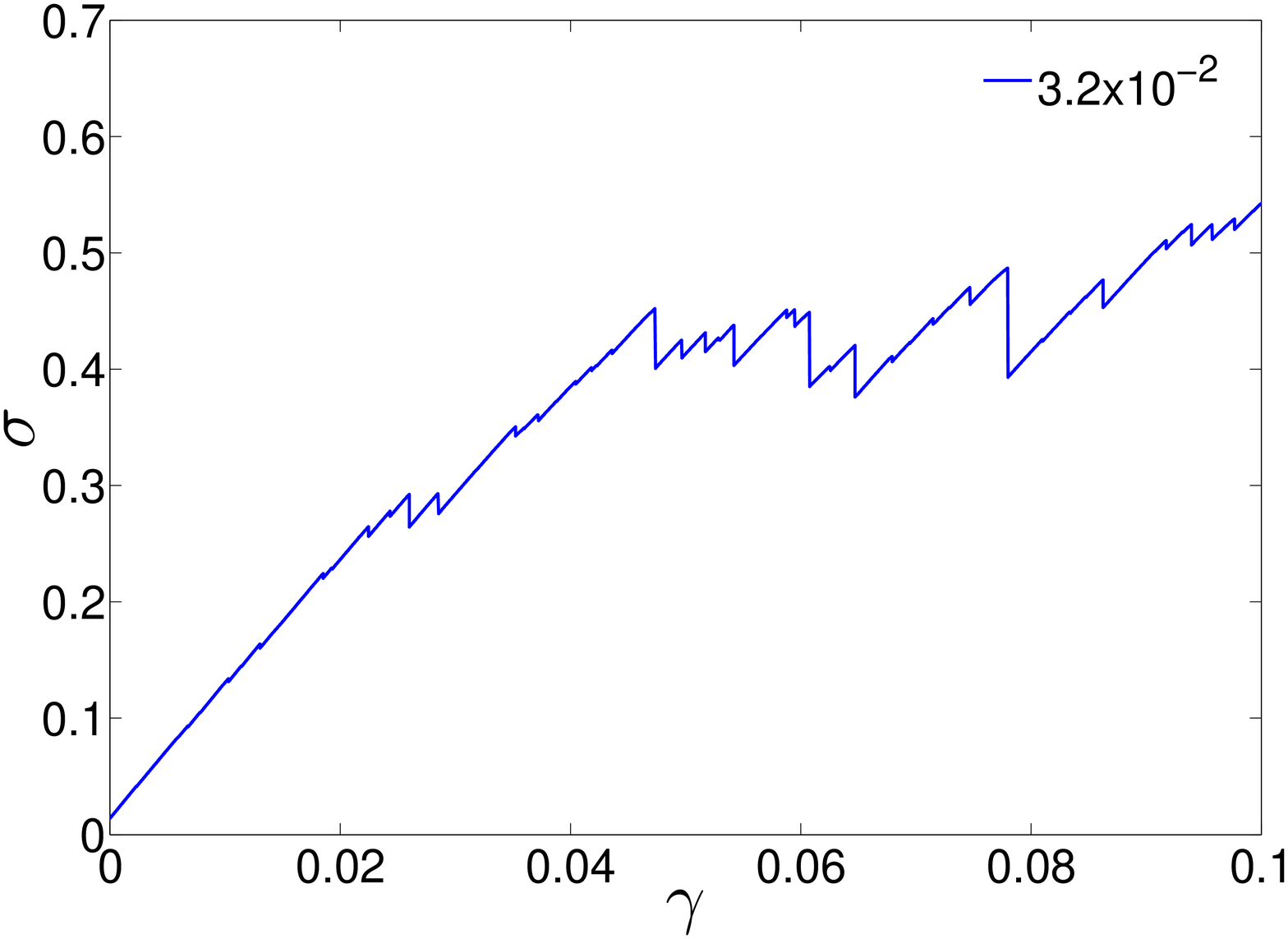}
\includegraphics[scale = 0.2]{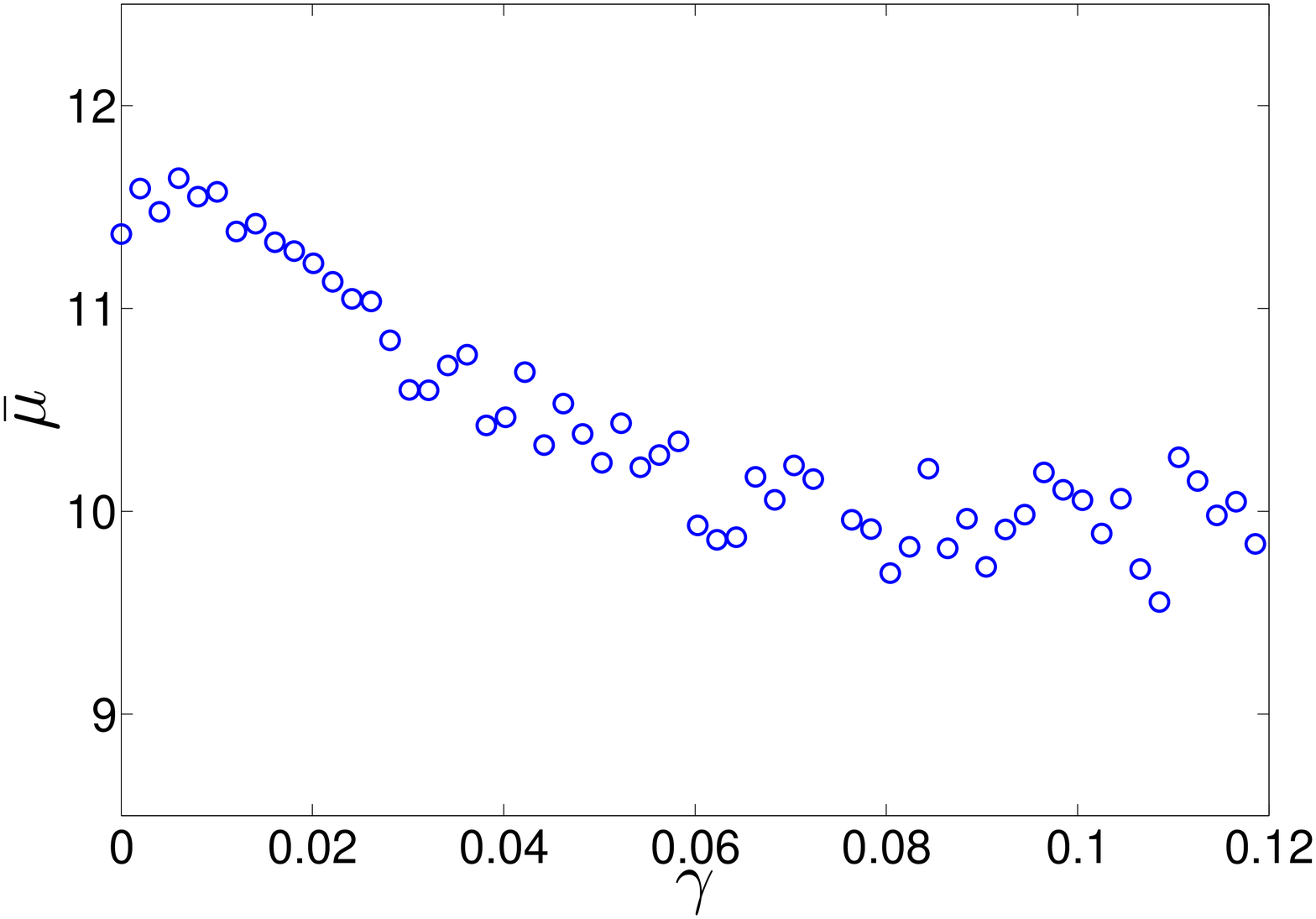}
\caption{(color online). Upper panel: Typical stress vs. strain in a quasi-static shear strain of our model amorphous solid, in the present case quenched from the supercooled liquid at a rate of $3.2\times 10^{-2}$ (in Lennard-Jones units, see appendix). We measure the shear modulus $\mu\equiv d\sigma/d\gamma$ in the piece-wise reversible straining steps between plastic drops.
Lower panel: The measured value of the shear modulus in the reversible steps between the plastic drops as a function of the strain $\gamma$. The shear modulus drops from a maximum value of the virgin material to as steady-state value at higher values of $\gamma$. The shear moduli are averaged over 25 different realizations, and are therefore denoted as $\bar \mu$. }
\label{sigvsgam}
\end{figure}

Our starting point is a discovery that occurred in our numerical simulations of binary  amorphous solids. The composition and preparation of the amorphous glasses by quenching them from the high temperature fluid is standard, and the necessary details are presented in the appendix. Once having created the samples, we strain them quasi-statically in athermal conditions using standard protocols (see appendix). A typical realization of the stress $\sigma$ vs. strain $\gamma$ is shown in the upper panel of Fig.~\ref{sigvsgam}. The stress increases linearly with the strain in a piece wise manner, punctuated by plastic drops. Between the plastic drops the shear modulus
$\mu \equiv d\sigma/d\gamma$ is almost constant up to the plastic drop where it becomes singular. The discovery is that the actual value of the piece-wise constant shear modulus decreases as the strain increases, until it reaches a steady state at higher values of $\gamma$. This is demonstrated in the lower panel of Fig.~\ref{sigvsgam}. The aim of this Letter is to offer an explanation of this phenomenon. In larger systems this phenomenon is harder to observe; in the thermodynamic limit the distance $\delta \gamma$ and the size of the stress drops shrink so much
that the stress vs. strain curve appears smooth \cite{09LP,10KLP}. It is then tempting to describe the phenomenon with some version of a series expansion \cite{11HKLP},
$
\sigma(\gamma)= \mu \gamma +B_2 \gamma^2 + B_3 \gamma^3+\dots
$
disregarding plasticity.
We argue that this is untenable. First, the radius of convergence of such a series is the nearest singularity i.e. the closest plastic drop  \cite{11HKLP}. Second, the system remains essentially linear between the plastic drops; we just need a reasonable model for the reducing shear modulus due to the effect of the accumulation of plastic drops.

We will explain the phenomenon by realizing that each one of the plastic drops localizes on some $p$ particles as shown below. We propose that once a region participates in a plastic drop, its local shear modulus will change from the virgin value $\mu_\ell$ to a smaller value $\mu_s$ which is computed below. The average shear modulus will be shown to be determined by a linear interpolation between the failed regions and the virgin regions that did not participate in a plastic drop. We will show below that this `two-state model' is surprisingly sufficient to explain the shear softening discussed here.

To make these remarks quantitative we recall that each plastic drop occurs when an eigenvalue of the Hessian matrix of the material hits zero \cite{12DKP}. The Hessian matrix $H_{ij}$ is defined by \cite{10KLLP}
\begin{equation}
H_{ij} \equiv \frac{\partial^2 U(\B r_1, \B r_2, \cdots \B r_N)}{\partial \B r_i \partial \B r_j}\ ,
\end{equation}
where $U(\B r_1, \B r_2, \cdots \B r_N)$ is the total energy of the system as a function of the positions of the particles $\{\B r_i\}$, $i=1,2,\cdots,N$. Between the plastic drops all the eigenvalues of $\B H$ are positive, and the system is mechanically stable. Upon approaching the plastic drop one of the eigenvalues of $\B H$, say $\lambda_P$, approaches zero following a square-root singularity as the system undergoes a saddle node bifurcation \cite{09LP,ML,12DKP}. At the same value of $\gamma$ where this happens, say $\gamma_p$, the ``plastic" eigenfunction associated with $\lambda_p$, say $\Psi_p(\B r_1,\B r_2, \cdots \B r_N)$, localizes on a typical quadrupolar structure.  This quadrupolar event identifies with the non-affine displacement associated with the local stress and energy release which consists the plastic event. We can determine the effective number of particles participating with the plastic event by measuring the {\bf participation ratio} $P$, which is defined as \cite{14GJMPS}
$
P(\B \Psi_p) \equiv \Big[\sum_{j=1}^N \B |\B \Psi_p^{(j)}|^4\Big]^{-1}, $
where $\B \Psi_p^{(j)}$ is the plastic eigenfunction projected on the $j$th particle. For fully extended modes
this number is of $O(N)$ whereas for localized modes it can be much smaller.

\begin{figure}
\includegraphics[scale = 0.16]{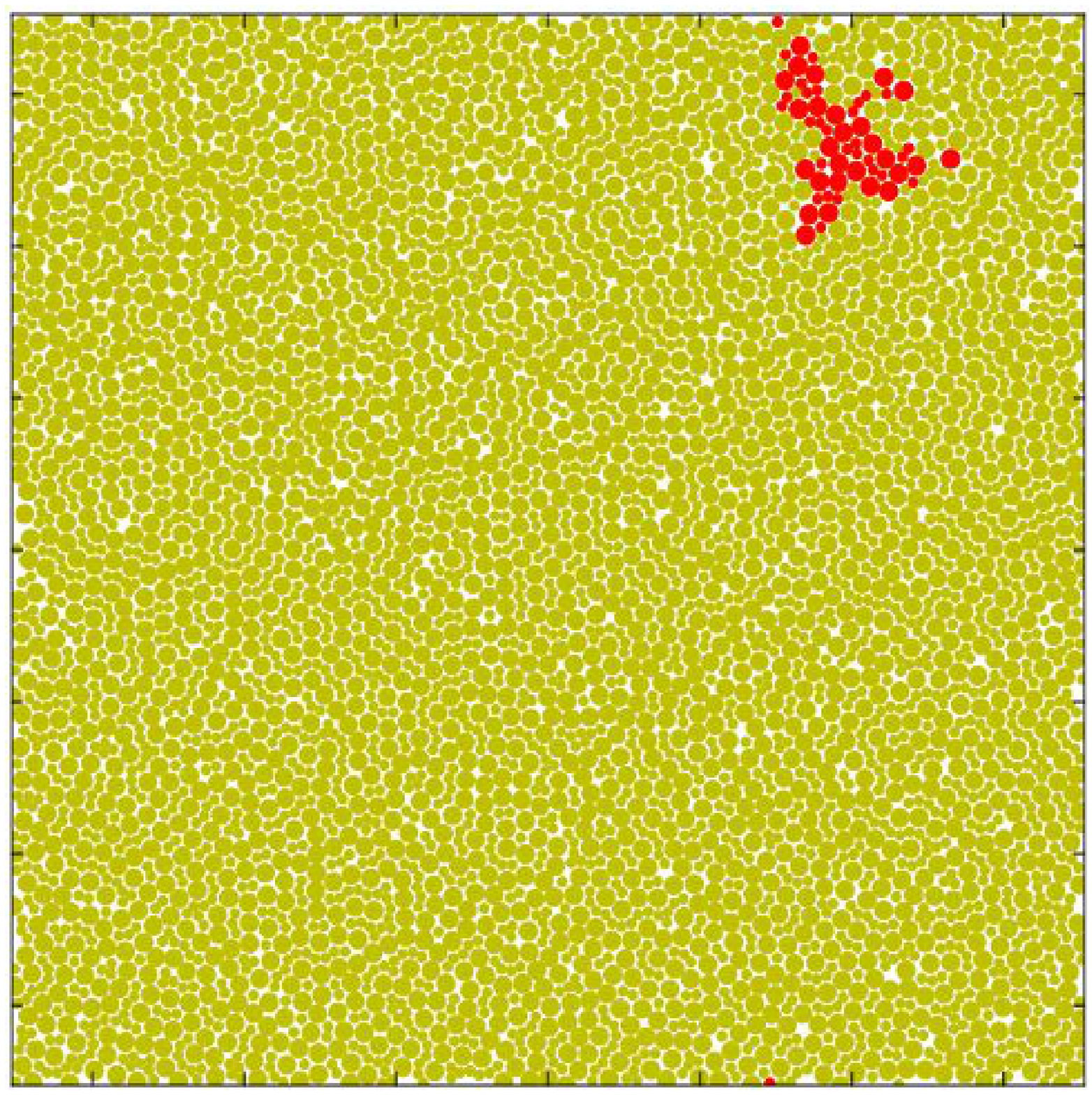}
\includegraphics[scale = 0.16]{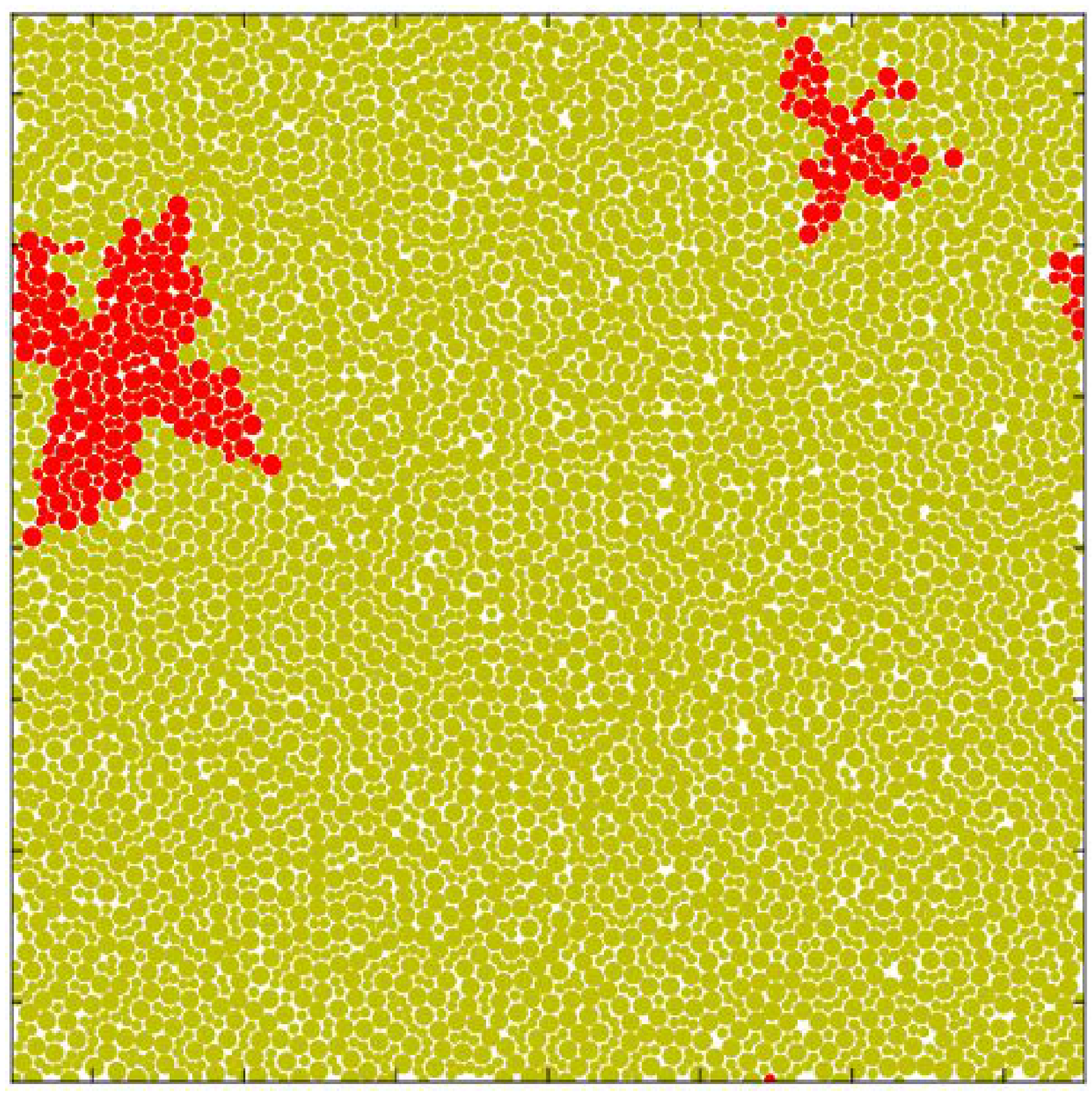}
\quad
\includegraphics[scale = 0.16]{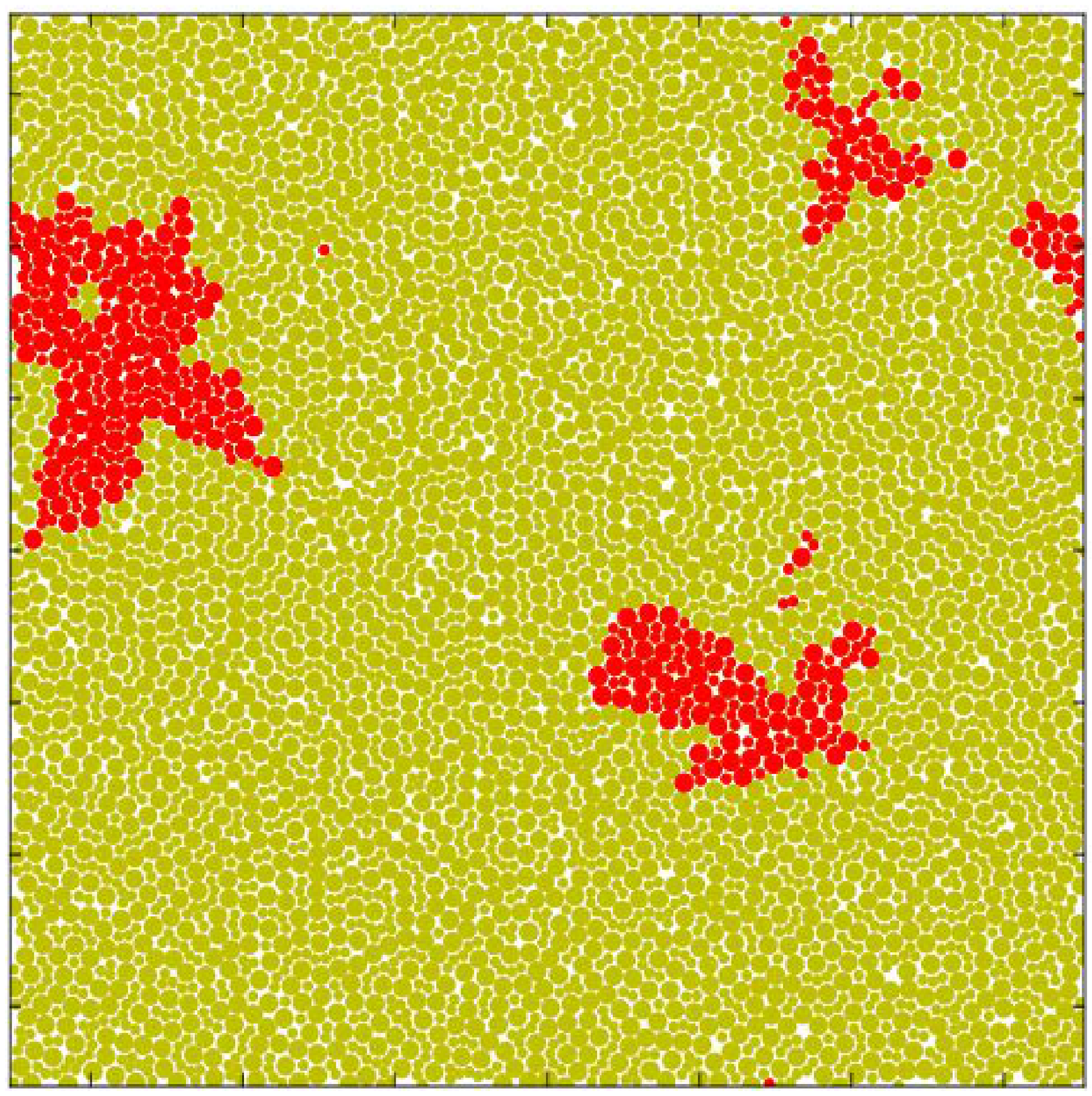}
\includegraphics[scale = 0.16]{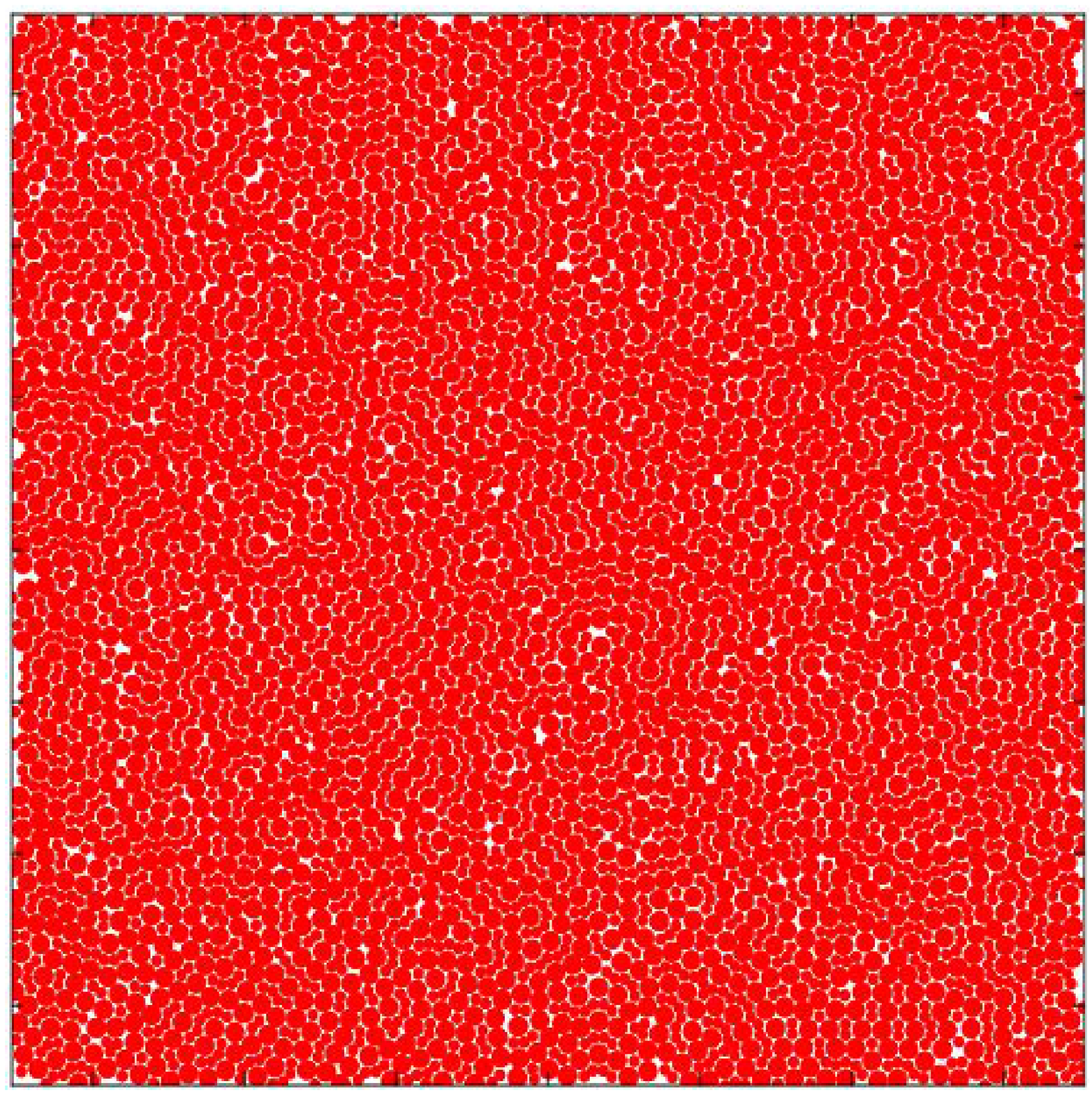}
\caption{(color online). Panel a): the first $P_1$ particles that participated in the first plastic event occurring at $\gamma=0.26\%$. Panel b) $p_1$ and $p_2$ particles, the second group participated in the second plastic event at $\gamma=0.6\%$. Panel c) $P_1, P_2$ and $P_3$ highlighted at $\gamma=1\%$. Panel d) At large values of $\gamma$ all the particles in the system participated in plastic events at some earlier value of $\gamma$.}
\label{red}
\end{figure}

Since we have many plastic events as the strain increases, and in each one the scenario just described repeats, we shall label the consecutive participation ratios as $P_1, P_2\dots$ for the first, second etc. plastic events. In Fig. \ref{red} we highlight the $P_1$ particles on which the first plastic event localized, and in the second and third panels we add the $P_2$ and $P_3$ particles on which the second and third plastic events localized. Obviously, as the strain increases, more and more regions get highlighted until eventually the whole system will be highlighted, see the last panel in Fig. \ref{red}.

To compute the two asymptotic values of the shear modulus in the virgin and fully plasticized sample we use the exact theory that prescribes the value of shear moduli as \cite{10KLLP}:
\begin{equation}
\mu=\frac{1}{V}\frac{\partial^2 U(\B r_1,\cdots,\B r_N;\gamma)}{\partial \gamma^2}-\frac{1}{V}
\B \Xi\cdot {\B H}^{-1} \cdot \B \Xi \ , \label{defmu}
\end{equation}
where the first term is the well known Born contribution. The second term exists only due to the non-affine
displacement $\B u_i$ (see appendix) and it includes the non affine ``force" $\B \Xi$ \cite{ML} defined as
$
\Xi_i \equiv \partial^2 U(\B r_1,\cdots,\B r_N)/\partial \B r_i \partial \gamma \ .
$
Using this exact formalism in the pure virgin state and in the fully failed state we come up with the value $\mu_\ell$ of the virgin material and $\mu_s$ of the fully plasticized counterpart. Averaging over realizations we obtain two averaged values
$\bar \mu_\ell$ and $\bar\mu_s$.

\begin{figure}
\includegraphics[scale = 0.20]{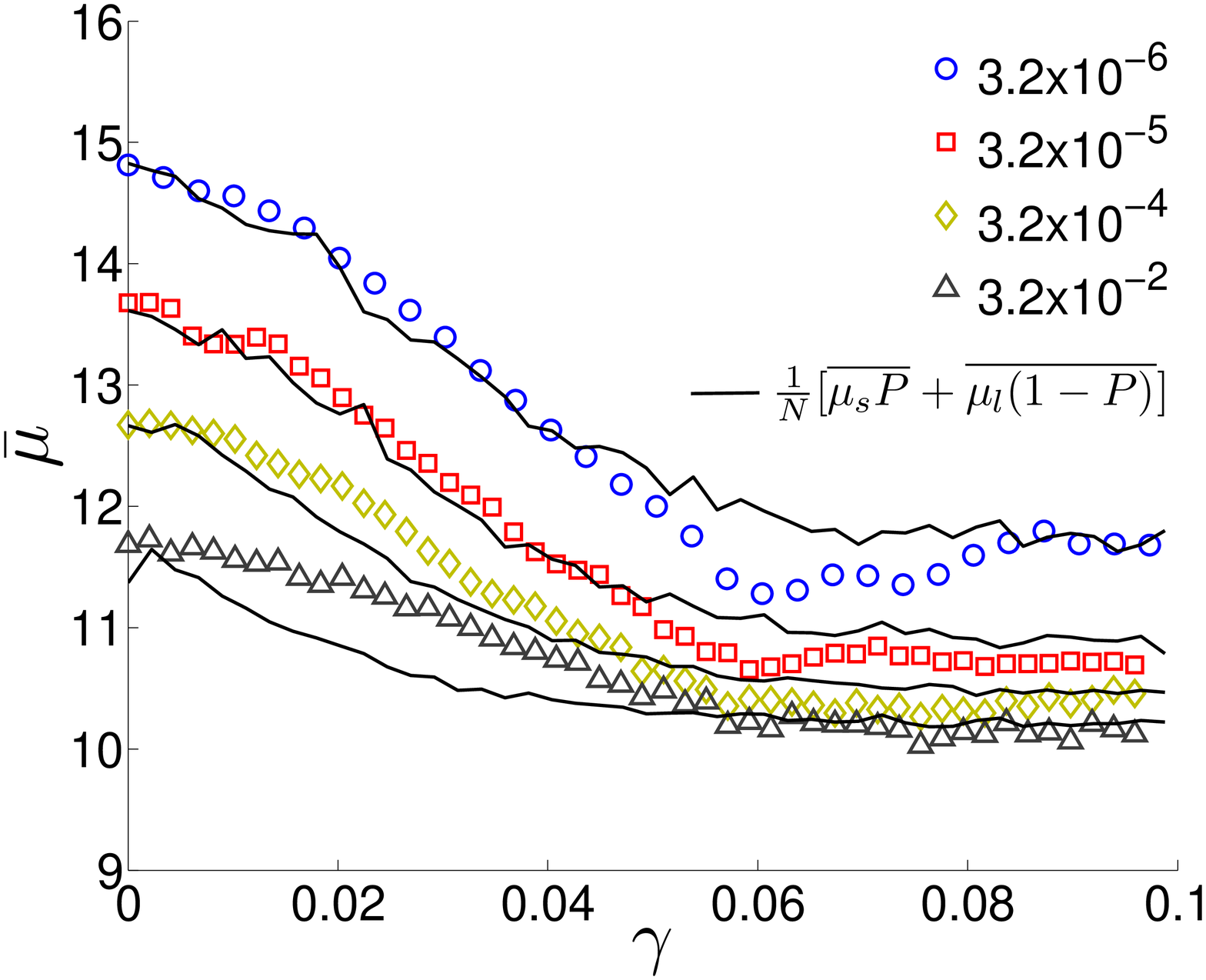}
\caption{The shear modulus $\bar \mu$ averaged over 25 realizations quenched at four different rates
from $3.2\times 10^{-2}$ to $3.2\times 10^{-6}$. In dots we show simulation data and in continuous line
the prediction of Eq. (\ref{barmu}).}
\label{interp}
\end{figure}

The first important result shown here is that the measured average shear modulus
$\bar \mu$ can be estimated as a linear interpolation between $\bar \mu_\ell$ and $\bar\mu_s$. For a given value of
$\gamma$ each realization experienced $k(\gamma)$ plastic events with participation ratios $P_1, P_2, \dots, P_{k(\gamma)}$.
We denote by $P$ the total number of particles which already participated in at least one plastic event (we do not double
count particles participating in more than one event). We then write
\begin{equation}
\bar \mu = \frac{1}{N}\left[\overline{P \mu_s} +\overline{(1-P) \mu_\ell}\right] \ ,
\label{barmu}
\end{equation}
where an overline stands for an average over realizations.
The quality of this model is demonstrated in Fig.~\ref{interp}. We conclude that a ``two state" model suffices to provide a surprisingly good model of the shear softening.
\begin{figure}
\includegraphics[scale = 0.20]{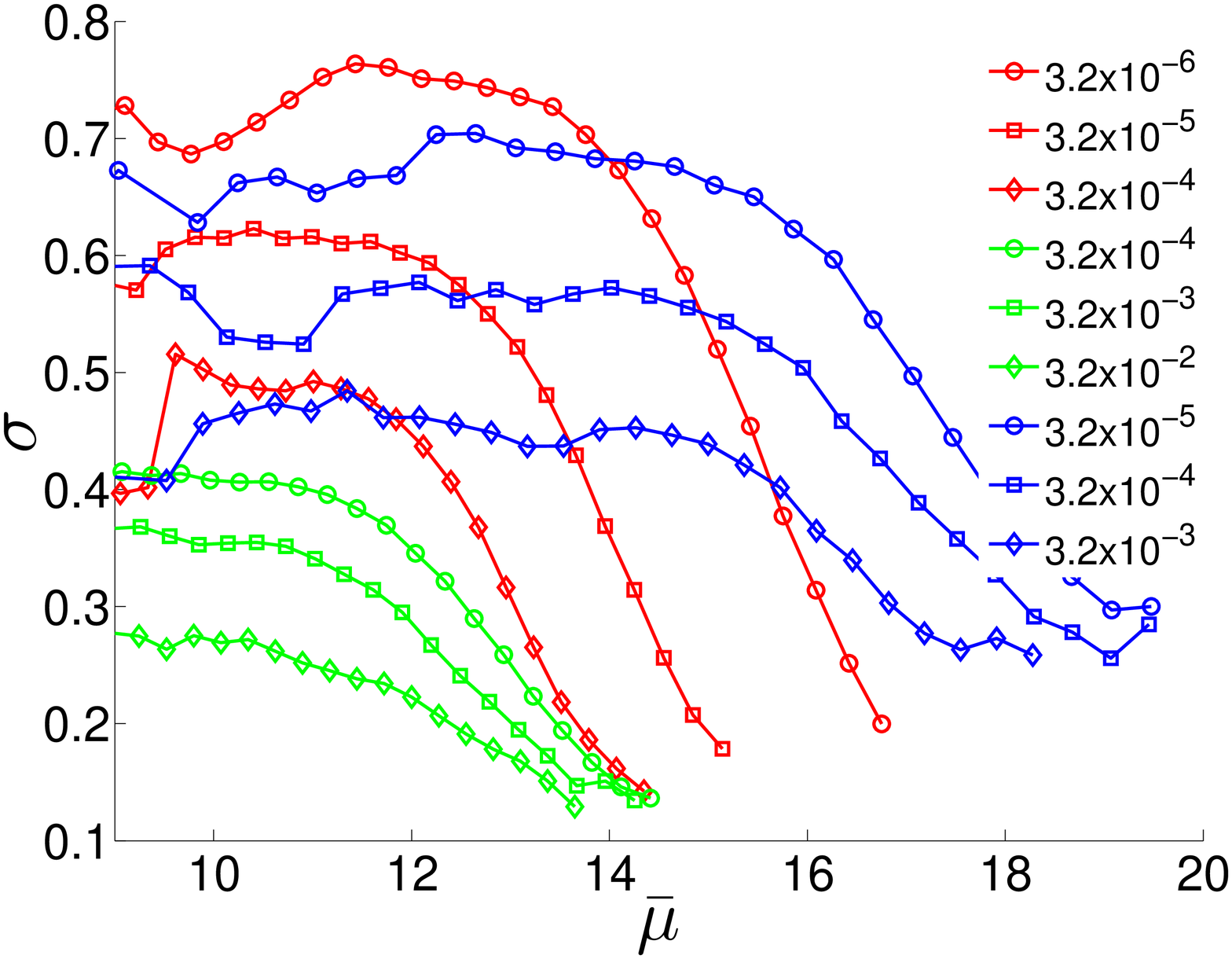}
\includegraphics[scale = 0.20]{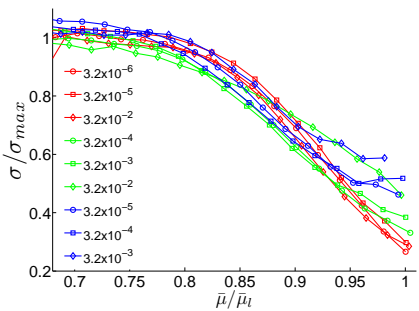}
\caption{(Color online). Upper panel: The stress vs. the piece-wise value of $\bar \mu$ for systems with 3 different potentials (in different colors, see appendix) for different quench rates (different symbols). Lower panel: the same data but with the stress normalized by its maximum value and $\bar \mu$ normalized by $\mu_\ell$. The data collapse indicates the existence
of a geometric interpretation. }
\label{quench}
\end{figure}

As is well known when the quench rate
from the supercooled liquid to the amorphous solid varies, so does the shear modulus of the virgin material $\mu_\ell$.
As a result, the whole stress-strain curce will depend on the quench rate of the virgin material. This variability is shown in the upper panel of Fig. \ref{quench} for three quench rates and also for three different interparticle potentials, and see appendix for details. However, it is easy to see that by rescaling $\sigma(\gamma)$ by its maximum value along the strain-stress curve, and by rescaling $\mu(\gamma)$ by $\mu_\ell$ we can collapse all the data
in the upper panel of Fig. \ref{quench} to one curve, see the lower panel of the same figure. This indicates strongly
that one should seek a simple geometric interpretation to the softening phenomenon.

\begin{figure}
\includegraphics[scale = 0.18]{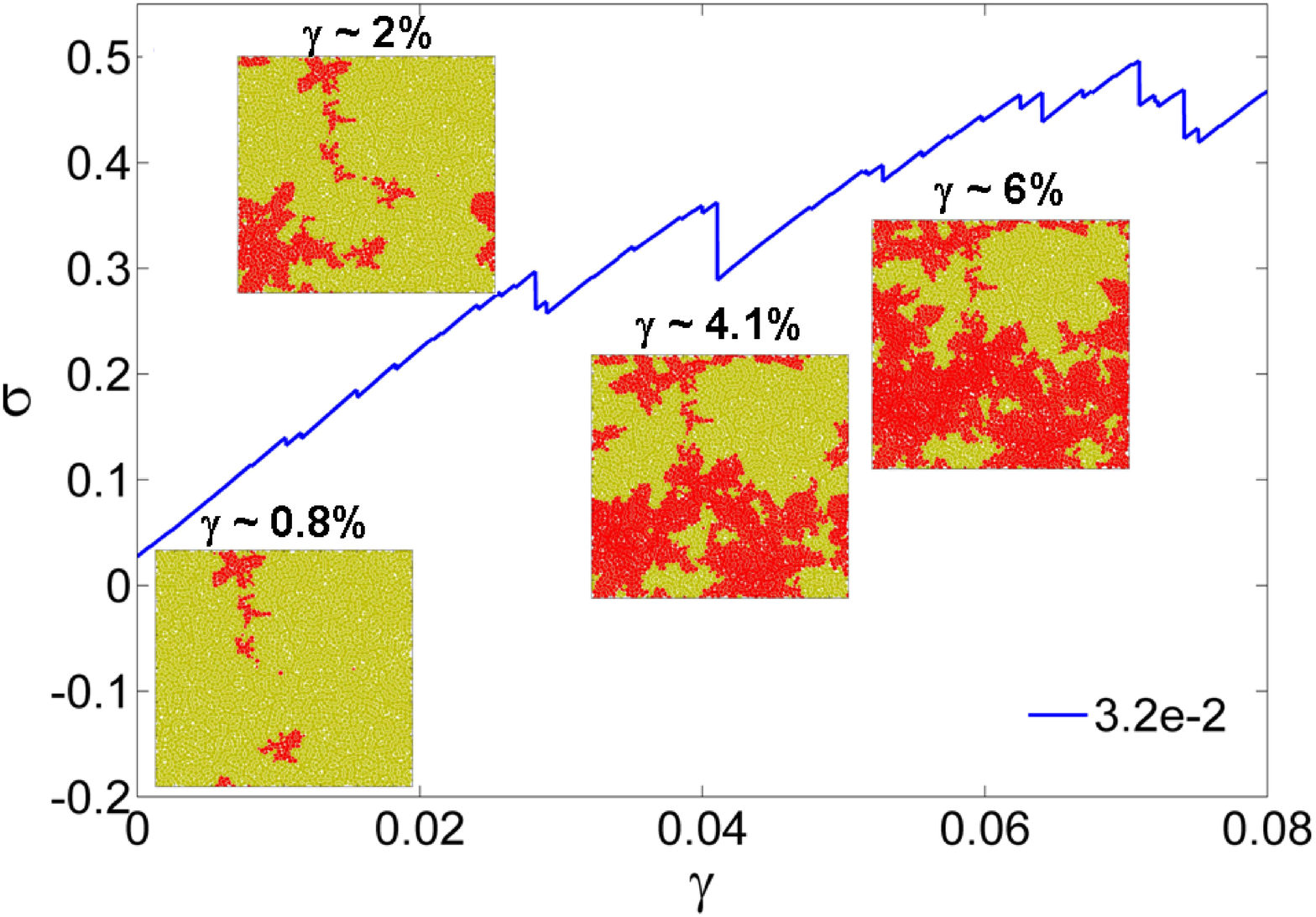}
\includegraphics[scale = 0.18]{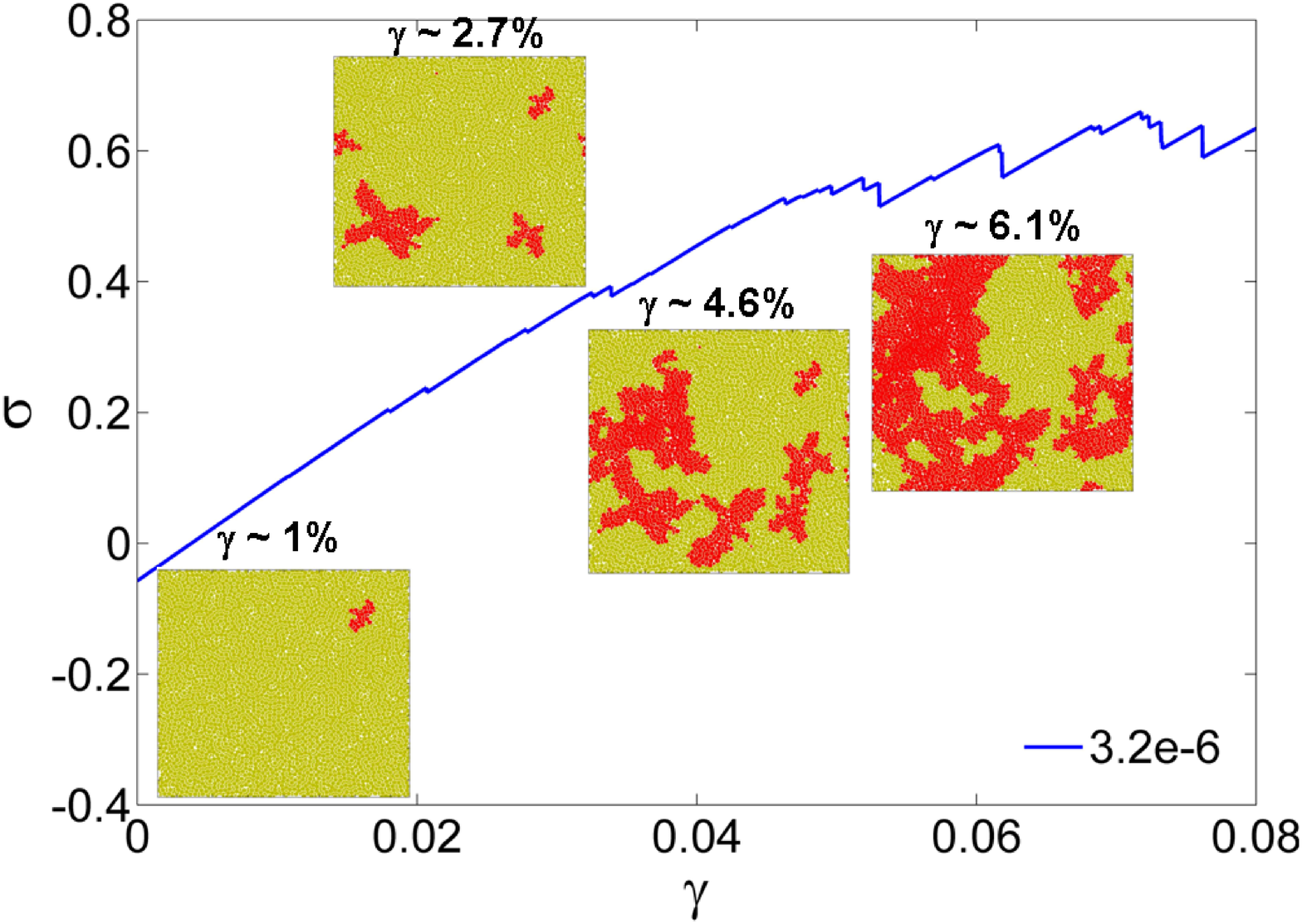}
\caption{Typical stress-strain curves for two quench rates ($3.2\times 10^{-2}$ and $3.2\times 10^{-6}$ respectively (for the first potential in the appendix), accompanied by the snapshots of the strained systems with particles that participated in plastic events highlighted in
red. The percolation of failed regions is always the beginning of sub-extensive stress drops.}
\label{beauty}
\end{figure}

Indeed, examining the increase in number of regions that had participated in plastic events we realize that the approach to the steady-state behavior is strongly correlated with a percolation of the plasticized regions across the system \cite{05SF}. This is demonstrated in Fig. \ref{beauty} in which we show the stress vs. strain curves of the fastest and slowest quenched samples together with the highlighted particles that participated in a plastic event. Note that yielding can appear either as a shear band for the slowly quenched system, or as a gradual yielding for the fastest quenched system, but in both cases the yield phenomenon is characterized by a percolation of the failed regions. Note that also after failure and the attainment of statistical steady state, the system is NOT a fluid.
It still has a finite shear modulus between the plastic drops, and it will not flow without further increase in strain. (A fluid of course flows with any infinitesimal stress).
The upshot of these observations is that the stress vs. strain curve will approach its steady-state behavior when the
failed part percolate throughout the system. This will be independent of the quench rate, explaining why the normalized
data in Fig.~\ref{quench} collapses for different potentials and different quench rates. The attainment of the steady state
is determined by the purely geometric accumulation of failed regions to a percolated cluster.

In summary, we examined carefully the phenomenon of shear softening and showed that a very simple ``two-state" model suffices to characterize it. Clearly, this simple model is not a complete theory and it cannot be expected to fit perfectly in
the percolation region. Indeed, the simple interpolation formula \ref{barmu} is less accurate precisely near yielding which is associated with the percolation.  Nevertheless the geometric approach has merits: we can collapse data for different amorphous solids that are characterized by different potentials and
are prepared with quench rates spanning four orders of magnitude. This surprising simplicity stems from the fact that for
every system there exists two extreme values of shear modulus, that of the virgin material and that of the failed material.
Extrapolating between them yields the measured piece-wise constant shear modulus. One could not hope for a simpler model. Of course it should be noted that the existence of this scenario is most apparent in systems of finite size. When the thermodynamic limit is approached the distance $\delta \gamma$ between plastic drops and their magnitude decrease with the
system size increasing. Thus one observes a misleadingly smoother curve, tempting one to develop a nonlinear elastic
theory disregarding the plastic drops. It is a sad fact however that a Taylor expansion of stress as a function of strain for athermal amorphous solids has a radius of convergence until the first plastic drop. This prevents the use of perturbative methods for any theoretical analysis of the non-linear regime from first principles. Our model predicts an average shear modulus that provides a surprisingly good fit to the simulation data. In addition, this mechanism is apparently generic and it leads to universal features of failure in amorphous solids.

{\bf Appendix}

 To prepare quality data for the present discussion we have employed a binary Lennard-Jones mixture with a potential energy for a pair of particles labeled  $i$ and $j$:
\begin{eqnarray}
&&U_{ij}(r_{ij}/\sigma_{ij}) =
 4\epsilon_{ij}\Big[\Big(\frac{\sigma_{ij}}{r_{ij}}\Big)^{12} - \Big(\frac{\sigma_{ij}}{r_{ij}}\Big)^{6}\Big] \ , \quad \text{for}~ \frac{r_{ij}}{\sigma_{ij}} \leq 1  \\
&&U_{ij}(r_{ij}/\sigma_{ij}) =\epsilon_{ij}\Big[A\Big(\frac{\sigma_{ij}}{r_{ij}}\Big)^{12} \!\!- \! B\Big(\frac{\sigma_{ij}}{r_{ij}}\Big)^{6} + C_0 + C_2\Big(\frac{r_{ij}}{\sigma_{ij}}\Big)^2\nonumber \\&&+ C_4\Big(\frac{r_{ij}}{\sigma_{ij}}\Big)^4 + C_6\Big(\frac{r_{ij}}{\sigma_{ij}}\Big)^6 \Big] \ ,~ \text{for}~1 < \frac{r_{ij}}{\sigma_{ij}} \leq 2.5 , \ . \nonumber
 \label{LJpot}
\end{eqnarray}
and $U(\frac{r_{ij}}{\sigma_{ij}})=0$ for $\frac{r_{ij}}{\sigma_{ij}} > 2.5$, where the parameters $A$ to $C_6$ are added to smooth the potential at a scaled cut-off of $r_{ij}/\sigma_{ij} = 2.5$ with two derivatives. The particles are labeled ``small"(S) or ``large"(L); the parameters $\sigma_{_{SS}}$, $\sigma_{_{LL}}$ and $\sigma_{_{LS}}$ were chosen as $2\sin(\pi/10)$, $2\sin(\pi/5)$ and $1$ respectively; $\epsilon_{_{SS}} = \epsilon_{_{LL}} = 0.5, \epsilon_{_{LS}} = 1$(see \cite{98FL}). The particle masses are all unity. All distances $|r_i-r_j|$ are normalized by $r_{SL}$. The energy is normalized by $\epsilon_{SL}$. Temperature was measured in units of $\epsilon_{SL}/k_B$ where $k_B$ is Boltzmann's constant. The number of particles in our simulations is varying between $4900$ to $10000$ at a number density $n = 0.985$ with a particle ratio $N_L/N_S = (1 + \sqrt{5})/4$. The mode coupling temperature $T_{MCT}$ for this system is known to be $0.325$. Time is normalized to $t_0 = \sqrt{\sigma_{LS}^{2}/\epsilon_{LS}}$. To prepare the glasses, we first start from a well equilibrated liquid at a high temperature $T = 1.2$, which is supercooled to $T = 0.35$ at a rate of $3.4\times10^{-3}$ using molecular dynamics. Secondly, we then equilibrate these supercooled liquids for times greater than $20\tau_{\alpha}$, where $\tau_{\alpha}$ is the time taken for the self intermediate scattering function to become $1\%$ of its initial value. Lastly, following this equilibration, we quench these supercooled liquids to a deep glassy phase at a temperature $T = 0.01$. This is done at various quench rates ranging from $3.2\times 10^{-2}$ to $3.2\times10^{-7}$ in jumps of one order of magnitude.

To prepare additional data for different potentials (mainly for the purpose of creating Fig. \ref{quench}) we repeated the
above procedure using two different potentials. For the Kob-Anderson model as a glass former we employed a $65:35$ binary mixture of Lennard-Jones particles at a total density of $\rho=\rho_L+\rho_S$, where the subscripts $L$ and $S$ refer to large and small particles, respectively. The particles interact through the potential given in Eq.\ref{LJpot}. The parameters $\epsilon_{ij}$, $\sigma_{ij}$ were chosen to agree with those in Ref.\cite{95KA}.

For the purely repulsive model as a glass former we employed again particles of two �sizes� but of equal mass of unity in two-dimensions, interacting via a pairwise potential of the form
\begin{equation}
U_{ij}(r_{ij}/\sigma_{ij}) =
 \epsilon\Big[\Big(\frac{\sigma_{ij}}{r_{ij}}\Big)^{k} + \sum_{l=0}^{q}c_{2l}\left(\frac{r_{ij}}{\sigma_{ij}}\right)^{2l}\Big] \ , \text{for} \frac{r_{ij}}{\sigma_{ij}} \leq x_c  \ ,
 \label{Repulsivepot}
\end{equation}
and zero otherwise. Here
$r_{ij}$ is the is the distance between particle $i$ and $j$, $\epsilon$ is the energy scale, and $x_c$ is the dimensionless length for which the potential will vanish continuously up to $q$ derivatives. The interaction lengthscale $\sigma_{ij}$ between any two particles $i$ and $j$ is $\sigma_{ij} = 1.0\sigma$, $\sigma_{ij} = 1.18\sigma$ and $\sigma_{ij} = 1.4\sigma$ for two 'small' particles, one 'large' and one 'small' particle and two 'large' particles, respectively. The coefficients $c_{2l}$ are given by
\begin{equation}
c_{2l}=\frac{(-1)^{l+1}}{(2q-2l)!!(2l)!!}\frac{(k+2q)!!}{(k-2)!!(k+2l)!!}x_c^{-(k+2l)}.
\end{equation}

We chose the parameters $x_c = 7/4$, $k = 10$ and $q = 6$. The unit of length $\sigma$ is set to be the interaction length scale of two small particles, and $\epsilon$ is the unit of energy. The density for all systems is set to be $N/V = 0.85\sigma^{-2}$.

We strain the glasses using an athermal quasistatic (AQS) protocol to examine their stress-strain curves. In each step of this procedure, the particle positions in the system are first changed by the affine
transformation,
\begin{equation}
x_i\rightarrow x_i + \delta\gamma y_i; \quad \quad y_i \rightarrow y_i
\end{equation}
This transformation takes the system away from mechanical equilibrium, so we therefore allow a second step, a nonaffine transformation $\B r_i \rightarrow \B r_i + \B u_i$ that annuls the forces between the particles, returning the system to mechanical equilibrium.

{\bf Acknowledgements}: this work was supported by the Israel Science Foundation, the German-Israeli Foundation and
by the ERC under the STANPAS ``ideas" grant.

\end{document}